\documentclass{article}
\input epsf

\begin{document}
\centerline{\bf Photometric Study of Kepler Asteroseismic Targets}

\vspace{0.3cm}
\centerline{by}

\vspace{0.3cm}
\centerline{J.~M o l e n d a - \.Z a k o w i c z$^1$, M.~J e r z y k i e w i c z$^1$ 
and A.~F r a s c a$^2$}

\vspace{0.3cm}
\centerline{$^1$ Astronomical Institute, University of  Wroc{\l }aw, Kopernika 11,}
\centerline{51-622 Wroc{\l }aw, Poland, e-mail: (molenda,mjerz)@astro.uni.wroc.pl}
\vspace{0.1cm}
\centerline{$^2$ Catania Astrophysical Observatory, Via S.Sofia 78,} 
\centerline{95123 Catania, Italy, e-mail: afr@oact.inaf.it}

\vspace{0.5cm}
\centerline{\it Received ...}

\vspace{1cm}
\centerline{ABSTRACT}

\vspace{0.5cm} 
{\small 

Reported are $UBV$ and $uvby \beta$ observations of 15 candidates for Kepler primary 
astero- seismic targets and 14 other stars in the Kepler field, carried out at the {\it 
M.G. Fracastoro\/} station of the Catania Astrophysical Observatory. These data serve to 
plot the 29 stars in two-parameter diagrams with the photometric indices (such as $B-V$ 
or $\delta m_1$) and the atmospheric parameters (such as the MK type or [Fe/H]) as 
coordinates. The two-parameter diagrams show no evidence of interstellar reddening. The 
photometric indices $B-V$ and $\beta$ are then used to derive photometric effective 
temperatures, $T_{\rm eff}(B-V)$ and $T_{\rm eff}(\beta)$. For  $T_{\rm eff}(B-V) > 
6400$ K, the photometric effective temperatures turn out to be systematically higher 
than spectroscopic effective temperatures by $311\,\pm\,34$ K and $346\,\pm\,91$ K for 
$T_{\rm eff}(B-V)$ and $T_{\rm eff}(\beta)$, respectively. For $T_{\rm eff}(B-V) < 6250$ 
K, the agreement between $T_{\rm eff}(B-V)$ and the spectroscopic effective temperatures 
is very good. The photometric surface gravities, derived from $c_1$ and $\beta$, show a 
range of about a factor of two greater than their spectroscopic counterparts do. 

}

\vspace{0.5cm}
\centerline{\bf 1. Introduction}

\vspace{0.5cm} 
This is a sequel to the spectroscopic study of Kepler\footnote{http://kepler.nasa.gov/} 
asteroseismic targets published by Molenda-\.Zakowicz et al.\ (2007, henceforth Paper 
I). In Paper I, we reported observations of 15 candidates for Kepler primary 
asteroseismic targets (PATS) and 14 other stars in the Kepler field, carried out at 
three observatories. For all these stars, we derived the radial velocities, effective 
temperatures, surface gravities, metallicities, and the projected rotational velocities 
from two separate sets of data by means of two independent methods. In addition, we 
estimated the MK type from one of these sets of data. 

In Paper I, three stars, HIP\,94335, HIP\,94734, and HIP\,94743, were found to have 
variable radial-velocity. For HIP\,94335 = FL Lyr, a well-known Algol-type eclipsing 
variable and a double-lined spectroscopic binary, the orbital elements computed from the 
new data agreed closely with those of Popper et al.\ (1986). For HIP\,94734 and 
HIP\,94743 = V\,2077 Cyg, discovered to be single-lined systems, orbital elements were 
derived for the first time. In addition, HIP\,94743 was demonstrated to be a detached 
eclipsing binary. 

In the present paper, the 15 PATS and 14 other stars in the Kepler field will be 
referred to collectively as the program stars. In Sect.\ 2, we give an account of our 
$UBV$ and $uvby \beta$ photoelectric photometry of these stars. In Sect.\ 3 we examine 
the distribution of the program stars in several two-parameter planes. In Sect.\ 3.1, we 
plot the program stars in the spectral type, $B-V$ plane using MK types from Paper I. 
The $UBV$ two-color diagram is presented in Sect.\ 3.2. We use both these diagrams to 
discuss the interstellar reddening of the program stars. The reddening is further 
investigated in Sect.\ 3.3 by means of comparing the $UBV$ ultraviolet excess with the 
metallicity parameter [Fe/H] from Paper I. In paying so much attention to the reddening 
of these not-so-distant stars we have been motivated by the fact that the Kepler Input 
Catalog, KIC-10, gives $E(B-V)$ as high as $0.06$ mag for some of our program stars. 
Sect.\ 3.4 and 3.5 are limited to program stars having spectral type G2 or earlier; in 
Sect.\ 3.4 we consider the $\beta$, $b-y$ diagram, and in Sect.\ 3.5 we examine 
correlations of the Crawford's metal-abundance parameters $\delta m_1(\beta)$ and 
$\delta m_1(b-y)$ with [Fe/H]. In Sect.\ 4 we derive effective temperatures of the 
program stars from their color indices and compare them with the spectroscopic effective 
temperatures from Paper I. Sect.\ 5 is again limited to program stars having spectral 
type G2 or earlier; for these stars we determine the surface gravities from the $uvby 
\beta$ indices and confront them with the spectroscopic values from Paper I. A summary 
is provided in Sect.\ 6. Details of the photometric reductions are given in the 
Appendix. 

\vspace{0.5cm}
\centerline{\bf 2. Observations and Results}

\vspace{0.5cm} 
The observations were obtained by JM-\.Z at the {\it M.G. Fracastoro\/} station (Serra La 
Nave, Mount Etna, elevation 1750 m) of the Catania Astrophysical Observatory on nine 
consecutive nights, from 2006 June 23 to July 1, with a photon-counting single-channel 
photometer, mounted in the Cassegrain focus of a 91-cm telescope. The photometer housed 
an EMI 9893QA/350 photomultiplier tube, cooled to $-$15$^o$~C. Johnson $UBV$ filters were 
used on four nights, from June 23 to 25, and on July 1. On six nights, from June 25 to 
30, the observations were taken with Str\"{o}mgren $uvby$ filters; on four nights, June 
25 to 28, H$\beta$ filters were also used. The $\beta$ photometry was not obtained for 
stars having spectral types later than G2. The raw magnitudes and color indices of the 
program stars were corrected for the effects of atmospheric extinction and transformed to 
the standard systems using standard stars, observed on the same nights as the program 
stars. Details of the photometric reductions are given in the Appendix.

The results of the $UBV$ and $uvby \beta$ photometry are given in Tables 1 and 2, 
respectively. The values listed are straight means of the magnitudes and the color 
indices obtained from individual observations. In Table 1, the number of observations, 
$N$, is greater for the magnitudes than for the color indices because the magnitudes 
were derived not only from observations with filter $V$, but also from observations with 
filter $y$. In case of the eclipsing binary HIP\,94335 = FL Lyr, the means were computed 
from observations outside eclipses. This resulted in reducing the number of individual 
$B-V$ and $U-B$ color-indices of HIP\,94335 to two. In case of HIP\,94743 = V2077 Cyg, 
all $UBV$ and $uvby \beta$ observations were obtained outside eclipses. However, we 
rejected two discrepant $UBV$ observations. This decreased the number of individual $V$ 
magnitudes to 12, and the number of individual $B-V$ and $U-B$ color-indices to two. The 
standard errors (s.e.) of the color indices of HIP\,94335 and HIP\,94743 given in Table 
1 are the medians of the standard errors of the remaining stars' color-indices. 

\begin{table}[t]
\begin{center}
\centerline{T a b l e \quad 1}
{UBV photometry of PATS and the remaining program stars}
\vspace{0.3cm}
{\small
\begin{tabular}         
{crcrrcrcr} \\
\hline\noalign{\smallskip}
HIP    & $V$\ \ \ & s.e. & N & $B-V$ & s.e. & $U-B$ & s.e. & N \\
\noalign{\smallskip}\hline\noalign{\smallskip}
\multicolumn{9}{c}{PATS}\\
\noalign{\smallskip}\hline\noalign{\smallskip}
 93011  &  9.641  & 0.008& 10&   0.401& 0.004& $-$0.047& 0.011&  4\\
 94335  &  9.316  & 0.004& 20&   0.554& 0.004&    0.003& 0.008&  2\\
 94497  &  9.921  & 0.006& 18&   0.933& 0.003&    0.673& 0.004&  8\\
 94565  &  9.351  & 0.004& 10&   0.527& 0.004&    0.058& 0.007&  4\\
 94734  &  9.489  & 0.007& 16&   0.641& 0.004&    0.135& 0.008&  4\\
 95098  &  9.501  & 0.005& 12&   0.537& 0.004& $-$0.003& 0.003&  6\\
 95637  &  9.182  & 0.006& 12&   0.373& 0.003& $-$0.088& 0.004&  6\\
 95733  & 11.022  & 0.009& 26&   0.752& 0.005&    0.185& 0.005& 10\\
 96634  &  9.152  & 0.010& 12&   0.807& 0.004&    0.398& 0.010&  4\\
 96735  &  9.200  & 0.015& 12&   0.871& 0.004&    0.526& 0.014&  4\\
 97219  &  9.028  & 0.021& 10&   0.829& 0.004&    0.468& 0.012&  4\\
 97337  & 11.041  & 0.007& 34&   1.259& 0.004&    1.078& 0.013& 18\\
 97657  &  9.531  & 0.004& 14&   1.066& 0.005&    0.966& 0.013&  8\\
 97974  & 10.017  & 0.010& 14&   0.625& 0.004& $-$0.003& 0.009&  6\\
 98655  & 10.495  & 0.008& 14&   0.767& 0.005&    0.332& 0.007&  6\\
\noalign{\smallskip}\hline\noalign{\smallskip} 
\multicolumn{9}{c}{The remaining program stars}\\
\noalign{\smallskip}\hline\noalign{\smallskip}
 91128  &  9.909  & 0.005& 30&   1.396& 0.005&    1.185& 0.012& 14\\
 92922  &  9.183  & 0.006& 12&   0.769& 0.004&    0.345& 0.007&  6\\
 94145  &  8.992  & 0.007&  8&   0.242& 0.002&    0.022& 0.017&  4\\
 94704  & 11.168  & 0.006& 28&   0.654& 0.002& $-$0.095& 0.006&  8\\
 94743  &  9.142  & 0.012& 12&   0.335& 0.004& $-$0.035& 0.008&  4\\
 94898  &  9.499  & 0.008& 16&   0.792& 0.002&    0.407& 0.005&  4\\
 95631  &  9.132  & 0.010& 12&   0.779& 0.004&    0.379& 0.006&  4\\
 95638  & 10.581  & 0.010& 34&   0.697& 0.003&    0.182& 0.012&  8\\
 95843  &  9.260  & 0.009&  8&   0.418& 0.002& $-$0.093& 0.008&  4\\
 96146  &  9.089  & 0.011&  8&   0.467& 0.005& $-$0.006& 0.016&  4\\
 97168  & 10.411  & 0.007& 16&   0.618& 0.004&    0.012& 0.009&  6\\
 98381  &  9.916  & 0.006& 18&   1.081& 0.004&    0.998& 0.008&  8\\
 98829  &  9.698  & 0.006& 10&   0.712& 0.006&    0.099& 0.015&  6\\
 99267  & 10.140  & 0.010& 22&   0.468& 0.004& $-$0.312& 0.012&  6\\
\noalign{\smallskip}\hline 
\end{tabular}                            
}                               
\end{center}                    
\end{table}

\begin{table}[t]
\begin{center}
\centerline{T a b l e \quad 2}
{Four-color and $\beta$ photometry of PATS and the remaining program stars}
\vspace{0.3cm}
{\small
\begin{tabular}
{cccccccccr} \\
\hline\noalign{\smallskip}
HIP    & $b-y$\ \ & s.e. & $m_1$ & s.e. & $c_1$ & s.e.&$\beta$&s.e. & N \\
\noalign{\smallskip}\hline\noalign{\smallskip}
\multicolumn{9}{c}{PATS}\\
\noalign{\smallskip}\hline\noalign{\smallskip}
    93011 &  0.258 &  0.009 &  0.178 &  0.013  & 0.520 &  0.010 &2.690& 0.010&   6\\
    94335 &  0.365 &  0.004 &  0.169 &  0.006  & 0.361 &  0.007 &2.623& 0.005&  18\\
    94497 &  0.567 &  0.011 &  0.403 &  0.014  & 0.271 &  0.020 & --  &  --  &  10\\
    94565 &  0.345 &  0.006 &  0.163 &  0.013  & 0.455 &  0.017 &2.639& 0.009&   6\\
    94734 &  0.423 &  0.007 &  0.167 &  0.008  & 0.407 &  0.008 &2.608& 0.007&  12\\
    95098 &  0.330 &  0.009 &  0.186 &  0.014  & 0.412 &  0.020 &2.617& 0.009&   6\\
    95637 &  0.244 &  0.007 &  0.164 &  0.009  & 0.512 &  0.006 &2.702& 0.004&   6\\
    95733 &  0.461 &  0.008 &  0.253 &  0.012  & 0.248 &  0.013 & --  &  --  & 16\\
    96634 &  0.506 &  0.007 &  0.278 &  0.010  & 0.344 &  0.009 & --  &  --  &   8\\
    96735 &  0.545 &  0.004 &  0.340 &  0.008  & 0.310 &  0.008 & --  &  --  &   8\\
    97219 &  0.502 &  0.022 &  0.306 &  0.015  & 0.365 &  0.013 & --  &  --  &   6\\
    97337 &  0.777 &  0.014 &  0.649 &  0.021  & 0.131 &  0.030 & --  &  --  &  16\\
    97657 &  0.646 &  0.007 &  0.535 &  0.010  & 0.241 &  0.012 & --  &  --  &   6\\
    97974 &  0.423 &  0.008 &  0.168 &  0.013  & 0.295 &  0.008 &2.589& 0.008&   8\\
    98655 &  0.473 &  0.013 &  0.284 &  0.018  & 0.293 &  0.011 & --  &  --  &   8\\
\noalign{\smallskip}\hline\noalign{\smallskip}                                
\multicolumn{9}{c}{The remaining program stars}\\                             
\noalign{\smallskip}\hline\noalign{\smallskip}                                
    91128 &  0.894 &  0.009 &  0.528 &  0.015  & 0.171 &  0.020 & --  &  --  &  16\\
    92922 &  0.527 &  0.009 &  0.182 &  0.014  & 0.451 &  0.015 & --  &  --  &   6\\
    94145 &  0.127 &  0.012 &  0.197 &  0.017  & 0.808 &  0.013 &2.808& 0.009&   4\\
    94704 &  0.420 &  0.007 &  0.186 &  0.010  & 0.106 &  0.015 & --  &  --  &  20\\
    94743 &  0.193 &  0.006 &  0.183 &  0.009  & 0.639 &  0.010 &2.740& 0.006&  10\\
    94898 &  0.514 &  0.007 &  0.236 &  0.010  & 0.395 &  0.007 & --  &  --  &  12\\
    95631 &  0.508 &  0.004 &  0.226 &  0.006  & 0.406 &  0.013 & --  &  --  &   8\\
    95638 &  0.442 &  0.009 &  0.205 &  0.010  & 0.344 &  0.015 & --  &  --  &  26\\
    95843 &  0.285 &  0.008 &  0.159 &  0.006  & 0.447 &  0.008 &2.629& 0.007&   4\\
    96146 &  0.323 &  0.006 &  0.156 &  0.013  & 0.479 &  0.018 &2.663& 0.009&   4\\
    97168 &  0.400 &  0.010 &  0.195 &  0.017  & 0.255 &  0.023 &2.566& 0.006&  10\\
    98381 &  0.643 &  0.008 &  0.561 &  0.016  & 0.234 &  0.012 & --  &  --  &  10\\
    98829 &  0.444 &  0.004 &  0.249 &  0.010  & 0.196 &  0.011 & --  &  --  &   4\\
    99267 &  0.372 &  0.007 &  0.077 &  0.009  & 0.166 &  0.013 &2.580& 0.008&  16\\
\noalign{\smallskip}\hline                                              
\end{tabular}                            
}                               
\end{center}
\end{table}

For the three binaries (the two mentioned in the preceding paragraph, and HIP\,94734) 
the combined magnitudes and color indices given in Tables 1 and 2 need duplicity 
corrections which will make the magnitudes and color indices  pertain to the brighter 
components. We shall derive the duplicity corrections presently, assuming these stars to 
be unreddened. 

According to Popper et al.\ (1986), the visual absolute magnitudes, $M_{\rm V}$, of the 
components of HIP\,94335 = FL Lyr are equal to 3.84 and 5.30 mag. Using the $M_{\rm V}$ 
difference, the MK type of F8\,V of the primary (see Paper I), and taking into account 
the fact that the secondary is a main-sequence star (see its $\log g$ in table XVII of 
Popper et al.\ 1986), we get G8\,V for the MK type of the secondary from the data 
tabulated by Lang (1992). The same result was obtained by Popper et al.\ (1986) from the 
$V-R$\, color-index. Using these MK types, the observed magnitude and color indices 
from Table 1, and the tables of intrinsic color-indices (Lang 1992), we derived the 
duplicity corrections for $V$, $B-V$, and $U-B$ (to be added to the quantity in 
question). They are listed in Table 3. 

For the remaining two stars, which are SB1 systems, the magnitude differences between 
components are not known. We have estimated the magnitude differences from the systems' 
mass functions (see Paper I) as follows. We begun with reading the mass of the primary 
from the MK type -- mass relation (Lang 1992). Next, for an assumed value of the orbital 
inclination, $i$, the mass function yielded the mass of the secondary. Using the same MK 
type -- mass relation as before and assuming the secondary to be a main-sequence star we 
obtained the secondary's spectral type. Then, the MK types of the components were 
transformed to $M_{\rm V}$ by means of the MK type -- $M_{\rm V}$ relation from Lang 
(1992). Finally, the duplicity corrections were obtained in the same way as for 
HIP\,94335 = FL Lyr. 

In case of HIP\,94743 = V2077 Cyg we assumed $i > 70^o$ because for smaller $i$ the 
eclipses would be rather unlikely to occur in this widely detached system. For $70^o < i 
< 90^o$, we obtained G7\,V for the MK type of the secondary, 1.76 mag for the $V$ 
magnitude difference between components, and the $UBV$ duplicity corrections listed in 
Table 3.

The last star, HIP\,94734, is not known to be eclipsing. Thus, the inclination of the 
orbit is unknown. We assumed $\sin i$ to be equal to $\pi/4$, the most probable value 
for a random distribution of the orbital inclinations. This yielded K7\,V for the MK 
type of the secondary, the $V$ magnitude difference equal to 3.40 mag, and the $UBV$ 
duplicity corrections listed in  Table 3.

\begin{table}[t]
\begin{center}
\centerline{T a b l e \quad 3}
\centerline{Duplicity corrections (in mag)}
\vspace{0.3cm}
{\small
\begin{tabular}
{crrr} \\
\hline\noalign{\smallskip}
for  & HIP\,94335 & HIP\,94734 & HIP\,94743 \\
\noalign{\smallskip}\hline\noalign{\smallskip}
 $V$   &    0.254\hspace{7pt} &    0.049\hspace{7pt} &  0.198\hspace*{7pt}\\
$B-V$  & $-$0.042\hspace{7pt} & $-$0.022\hspace{7pt} & $-$0.053\hspace*{7pt}\\
$U-B$  & $-$0.044\hspace{7pt} & $-$0.015\hspace{7pt} & $-$0.029\hspace*{7pt}\\
$b-y$  & $-$0.031\hspace{7pt} & $-$0.018\hspace{7pt} & $-$0.040\hspace*{7pt}\\
$m_1$  &    0.002\hspace{7pt} &    0.006\hspace{7pt} &    0.013\hspace*{7pt}\\
$c_1$  & $-$0.005\hspace{7pt} &    0.003\hspace{7pt} &   0.008\hspace*{7pt}\\
$\beta$&    0.018\hspace{7pt} &    0.003\hspace{7pt} &   0.028\hspace*{7pt}\\
\noalign{\smallskip}\hline 
\end{tabular}
}
\end{center}
\end{table}

For the $uvby \beta$ indices, we obtained the duplicity corrections using the 
above-derived magnitude differences between components; for $\beta$ we also used the 
unreddened relation between this index and $b-y$ from Crawford (1975a) and Olsen (1988) 
to estimate the $\beta$ indices of the secondaries. The duplicity corrections are given 
in Table 3. 

\vspace{0.5cm}
\centerline{\bf 3. Two-Parameter Diagrams}

\vspace{0.5cm} 
{\em 3.1. The $B-V$ color-indices and the MK types}

\vspace{0.5cm} 
In Fig.\ 1, the $B-V$ color-indices from Table 1 are plotted vs.\ the spectral type 
from Table 9 of Paper I. In case of the binaries, HIP\,94335, HIP\,94734 and HIP\,94743, 
we applied the duplicity corrections to $B-V$ (see the preceding section). The solid 
line represents the intrinsic relation for luminosity class (LC) V, and the dashed line, 
for LC III (Lang 1992). The distances, $r$ (in parsecs) used in coding the symbols, 
were obtained from the Hipparcos parallaxes (ESA 1997) revised by van Leeuwen (2007). 
The symbols joined with dotted lines are the two subdwarfs (open diamonds) and the two 
LC III stars (open circles). The remaining program stars were classified in Paper I to 
LC V, IV-V, IV or IV-III. 

As can be seen from Fig.\ 1, most program stars lie within less than one spectral 
subtype of the intrinsic relations. Five deviate by two subtypes. Three deviate by more 
than two subtypes. The latter are the two subdwarfs (the open diamonds joined with 
dotted line) and HIP\,94743, the small filled circle at $B-V \approx 0.28$ mag. Both 
subdwarfs deviate from the intrinsic relation for LC V by about four subtypes in the 
direction of earlier types. Qualitatively, this may be understood as an abundance 
effect: in F and G stars the lines of neutral metals strengthen with advancing type, so 
that smaller metallicity may mimic earlier type. In case of HIP\,94743 = V2077 Cyg, the 
color index indicates a spectral type of about A9, some three subtypes earlier than F2, 
assigned to it in Paper I. This discrepancy would disappear if the color index were {\em 
redder\/} by $0.07$ mag, but an error of this magnitude in the $B-V$ seems rather 
unlikely. This is supported by the fact that the star's $uvby \beta$ indices also 
indicate earlier spectral type (see Sect.\ 3.4). 

\begin{figure} 
\epsfbox{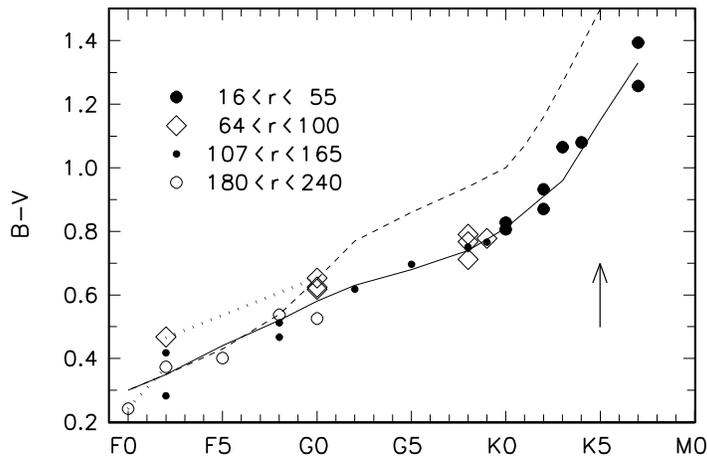}
\caption{The $B-V$ color-indices of the program stars plotted as a function of the 
spectral type. The solid line represents the intrinsic relation for luminosity class V, 
and the dashed line, for luminosity class III (Lang 1992). The distances, $r$ (in 
parsecs) used in coding the symbols, were obtained from the revised Hipparcos 
parallaxes (van Leeuwen 2007). The symbols joined with dotted lines are the two 
subdwarfs (open diamonds) and the two luminosity III stars (the leftmost open circles). 
The arrow is the reddening vector $E(B-V) = 0.2$ mag.} 
\end{figure}

In $B-V$, the points in Fig.\ 1 seem to scatter randomly around the LC V intrinsic 
relation. (Note that the two LC III program stars fall in the lower left-hand corner of 
the figure where the LC V and LC III lines coincide.) In order to quantify this 
impression, we computed the deviations in $B-V$ from the intrinsic relations, $(B-V) - 
(B-V)_0$. For LC V and LC III we read $(B-V)_0$ from the Lang's (1992) relations shown 
in the figure. For LC IV we used intrinsic values half-way between LC V and LC III, for 
LC IV-V, the values half-way between LC V and LC IV, and for LC III-IV, those half-way 
between LC IV and LC III. The mean deviation (omitting the two subdwarfs) 
turned out to be equal to $-0.009\,\pm\,0.010$ mag. In the intervals of distance 
indicated in the figure, the mean deviations amounted to $0.016\,\pm\,0.020$ mag for 
$16<r<55$ pc, $-0.026\,\pm\,0.023$ mag for $64<r<100$ pc, $-0.007\,\pm\,0.015$ mag for 
$107<r<165$ pc, and $-0.031\,\pm\,0.020$ mag for $180<r<240$ pc. 

Note that de-reddening makes the $(B-V) - (B-V)_0$ deviations decrease. Therefore, 
de-reddening will make the overall agreement with the intrinsic relations worse than that 
seen in the figure, except for an unlikely case of $E(B-V) \leq 0.030$ mag for $r < 55$ 
pc and $E(B-V) = 0$ mag for $r > 64$ pc. We conclude that there is no evidence in Fig.\ 1 
for $E(B-V) > 0.00$ mag even for the most distant stars in our sample. 

\vspace{0.5cm} 
{\em 3.2. The $UBV$ Two-Color Diagram}

\vspace{0.5cm} 
Fig.\ 2 shows the program stars plotted in the two-color diagram using the $B-V$ and 
$U-B$ color-indices from Table 1. In case of the binaries, HIP\,94335, HIP\,94734 and 
HIP\,94743, we applied the duplicity corrections derived in Sect.\ 2. The distances, $r$ 
(in parsecs) used in coding the symbols, were obtained from the revised Hipparcos 
parallaxes (van Leeuwen 2007). In the figure there is also plotted the Hyades 
main-sequence two-color line, explained in the next paragraph, and the Hyades line 
reddened with $E(B-V) = 0.05$ mag and $E(U-B) = 0.72 E(B-V)$. 

\begin{figure} 
\epsfbox{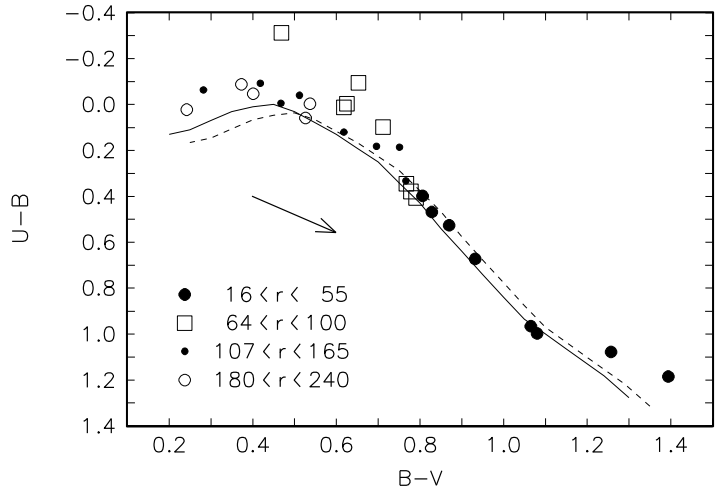}
\caption{Program stars in the two-color plane. The distances, $r$ (in parsecs) used in 
coding the symbols, were obtained from the revised Hipparcos parallaxes (van Leeuwen 
2007). The solid line represents the Hyades two-color relation.  The arrow is the 
reddening vector $E(B-V) = 0.2$ mag and $E(U-B) = 0.72 E(B-V)$. The dashed line is the 
Hyades relation reddened with $E(B-V) = 0.05$ mag.} 
\end{figure}

The two frequently used Hyades two-color relations are (1) that given by Sandage \& 
Eggen (1959) in their Table III, and (2) Johnson's (1966) main-sequence relation, listed 
in his Table II (used in lieu of the Hyades line by, e.g., Cameron 1985). The two lines 
agree closely for $0.20 < B-V < 0.95$ mag. For $B-V > 0.95$ mag, Johnson's (1966) line 
veers upwards. In order to decide which line is the more nearly correct one, we have 
used the list of the Hyades' members composed by Pinsonneault et al.\ (2004) and $UBV$ 
color-indices determined by Johnson and Knuckles (1955) and Mendoza (1967). Plotted in 
the $B-V$, $U-B$ plane, these data resulted in a point-diagram.  A free-hand curve drawn 
through the points coincided with the two-color relation of Sandage and Eggen (1959). 
The solid line in Fig.\ 2 is this relation, slightly smoothed on the red side of $B-V = 
1.00$ mag and extended to $B-V = 1.30$ mag, 0.20 mag beyond the last $B-V$ value of 
Sandage \& Eggen (1959). For $1.05 < B-V < 1.30$ mag, Johnson's (1966) main-sequence 
relation -- not shown in the figure to avoid crowding -- would coincide with our $E(B-V) 
= 0.05$ mag line. 

In Fig.\ 2, most stars with $B-V < 0.76$ mag show ultraviolet excess. This will be 
discussed in the next subsection. Stars with $0.76 < B-V < 1.10$ mag lie on or very 
close to the Hyades line. Assuming zero reddening for the Hyades (Crawford 1975a, Taylor 
1980), we conclude that these stars are unreddened. 

HIP\,97337 and HIP\,91128, the two red stars having $B-V > 1.25$ mag (both classified to 
K7\,V in Paper I), deviate from the Hyades line. If the deviations were caused by 
reddening, the $E(B-V)$ would have to be equal to 0.21 mag for HIP\,97337 and 0.34 mag 
for HIP\,91128. Neither value agrees with the stars' $(B-V) - (B-V)_0$ deviations in 
Fig.\ 1 and, of course, with the fact that they are among the least distant stars in our 
sample ($37\,\pm\,2$ and $15.5\,\pm\,0.2$ pc, respectively). 

\vspace{0.5cm} 
{\em 3.3. The Ultraviolet Excess and\, {\rm [Fe/H]}}

\vspace{0.5cm}
The ultraviolet excess, $\delta(U-B)$, is defined as the difference between the Hyades 
$U-B$ at the star's $B-V$ and the star's $U-B$. It is a function of the star's LC, 
metallicity, reddening and rotation. For the stars plotted in Fig.\ 2, the last factor 
will be neglected. Some justification for this comes from the fact that 22 stars in our 
sample have $v\sin i < 10$ km\,s$^{-1}$, and the remaining ones have $20 < v\sin i < 35$ 
km\,s$^{-1}$ (see Table 12 of Paper I). The first factor can be corrected for using the 
MK luminosity classification of Paper I and the intrinsic UBV color-indices for the LCs 
V and III tabulated by Lang (1992). The correction, to be subtracted from $\delta(U-B)$, 
amounts to the difference between the tabular $U-B$ at the star's $B-V$ and LC and the 
tabular $U-B$ at the same $B-V$ and LC V. For LC IV, we assumed the corrections to be 
equal to one-half of those for LC III. Likewise, the corrections for the LCs IV-V and 
III-IV were assumed to be equal to one-quarter and three-quarters of those for LC III, 
respectively. The corrections ranged from 0.010 mag for HIP\,94898 (G8\,IV) to 0.078 mag 
for HIP\,95637 (F2\,III). 

The $\delta(U-B)$ values, corrected in this way for the program stars of luminosity 
classes IV-V, IV, III-IV, and III, are listed in the third column of Table 4. These 
values are functions of metallicity and $B-V$, so that stars with the same metal 
abundance but different $B-V$ will have different $\delta(U-B)$. In the fourth column of 
Table 4 there are given the $\delta$ correction factors, $\delta 0.6/\delta$, which can 
be used to transform the observed $\delta (U-B)$ into $\delta (0.6)$, a measure of 
metallicity independent of $B-V$. We computed these factors from the $\delta = f(B-V)$ 
relation given by Sandage (1969) in his Table 1A. In case of two program stars that have 
$B-V < 0.35$ mag, HIP\,94145 and HIP\,94743, we had to extrapolate the $\delta = f(B-V)$ 
relation beyond $B-V = 0.35$ mag, the smallest $B-V$ in Sandage's (1969) Table 1A. 
Extrapolation was also needed for the subdwarf HIP\,99267 because its $\delta(U-B)$ is 
larger than the largest $\delta(U-B)$ in Sandage's (1969) Table 1A. These three 
$\delta$ correction factors are less secure than the other ones. For the two red stars, 
HIP\,97337 and HIP\,91129, no $\delta$ correction factors were derived. Note that most 
$\delta$ correction factors in Table 4 are equal to 1.000. 

\begin{figure} 
\epsfbox{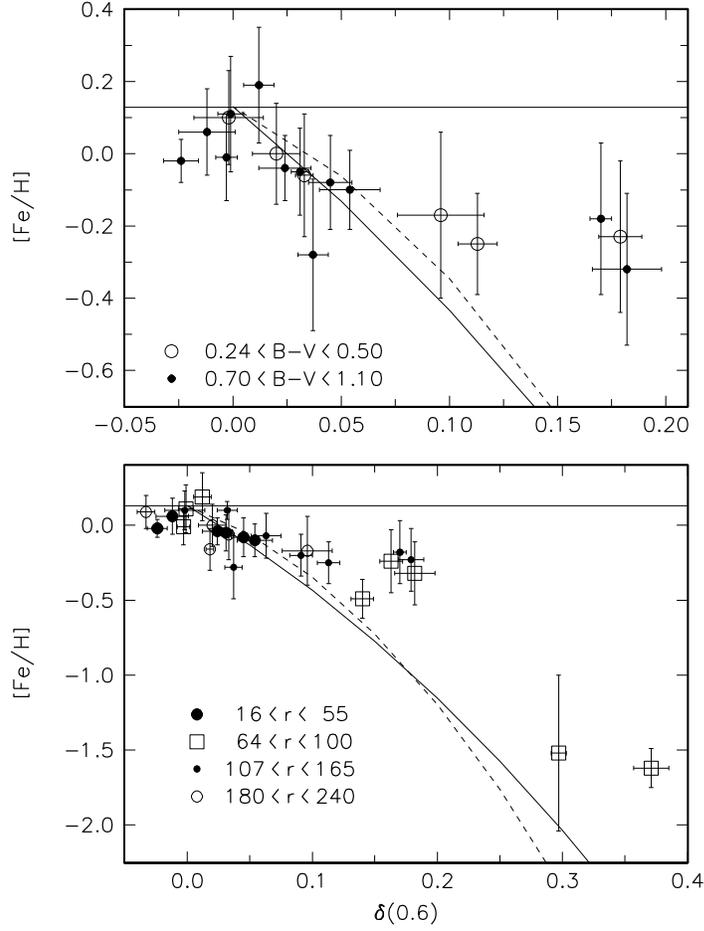}
\caption{The metallicity parameter [Fe/H] plotted as a function of $\delta(0.6)$. The 
lower panel contains all program stars except the two red ones, HIP\,97337 and HIP\,91128. 
In the upper panel, which shows the upper left-hand part of the lower panel, stars with 
$0.50 < B-V < 0.70$ are not plotted. Note that in the lower panel the symbols are coded 
as in Fig.\ 2, while in the upper panel the symbols are coded according to the $B-V$. The 
horizontal solid lines indicate [Fe/H] =  0.13, the Hyades' metallicity parameter 
according to Boesgaard \& Friel (1990). The inclined solid line represents the 
[Fe/H]--$\delta(0.6)$ relation for main-sequence stars due to Karata\c{s} \& Schuster (2006), 
and the dashed line, that due to Cameron (1985).} 
\end{figure}

In Fig.\ 3, the metallicity parameter, [Fe/H], is plotted as a function of $\delta(0.6)$. 
The standard errors of $\delta(0.6)$ were assumed to be equal to the standard 
error of the $U-B$ color-index (see Table 1) times the $\delta$ correction factor. In 
both panels, the horizontal solid line indicates [Fe/H] =  0.13, the Hyades' value of the 
metallicity parameter according to Boesgaard \& Friel (1990). The inclined lines 
represent the [Fe/H]--$\delta(0.6)$ relations for main-sequence stars due to Karata\c{s} \& 
Schuster (2006) and Cameron (1985). The lower panel contains all program stars except the 
two red ones, HIP\,97337 and HIP\,91128. In the upper panel, which shows the upper 
left-hand part of the lower panel, there are plotted stars having (1)~$B-V < 0.50$ mag, or 
(2)~$B-V > 0.70$ mag. As can be seen from Fig.\ 2, de-reddening group 1 stars would 
increase their ultraviolet excesses, while the reverse is true for group 2 stars. Thus, 
de-reddening these stars alters their position in relation to the [Fe/H]--$\delta(0.6)$ 
lines in Fig.\ 3. In the next paragraph, we shall use this facts to discuss the 
reddening. Stars with $0.50 < B-V < 0.70$ mag are not plotted in the upper panel of Fig.\ 
3 because in this $B-V$ interval the reddening vector is very nearly parallel to the 
Hyades line (see Fig.\ 2), and therefore $\delta (U-B)$ will not be affected by 
reddening. Consequently, the position of these points in relation to the 
[Fe/H]--$\delta(0.6)$ lines in Fig.\ 3 carries no information about $E(B-V)$. 

\begin{table} 
\begin{center}
\centerline{T a b l e \quad 4}
{The color index, ultraviolet excess, $\delta$ correction factor, and the metallicity 
parameter [Fe/H] from Paper I}
\vspace{0.3cm}
{\small
\begin{tabular}           
{cclccc} \\
\hline\noalign{\smallskip}
HIP    & $B-V$ & $\delta (U-B)$ & $\delta 0.6/\delta$& [Fe/H] & s.d. \\
\noalign{\smallskip}\hline\noalign{\smallskip}
\multicolumn{6}{c}{PATS}\\
\noalign{\smallskip}\hline\noalign{\smallskip}
 93011  & 0.401 &\hspace{4pt} 0.019*&1.035 &  $\ \ $0.00 & 0.14\\
 94335  & 0.512 &\hspace{4pt} 0.084 &1.078 &  $-$0.20 & 0.14\\
 94497  & 0.932 &\hspace{4pt} 0.031 &1.000 &  $-$0.05 & 0.12\\
 94565  & 0.526 &$-$0.033* &1.000 &$\ \ $0.09 & 0.11\\
 94734  & 0.618 &\hspace{4pt} 0.032 &1.000 &  $\ \ $0.10 & 0.06\\
 95098  & 0.537 &\hspace{4pt} 0.018* &1.000 &  $-$0.16 & 0.14\\
 95637  & 0.373 &\hspace{4pt} 0.031* &1.056 &  $-$0.06 & 0.17\\
 95733  & 0.751 &\hspace{4pt} 0.157 &1.080 &  $-$0.18 & 0.21\\
 96634  & 0.806 &\hspace{4pt} 0.045 &1.000 &  $-$0.08 & 0.13\\
 96735  & 0.870 &\hspace{4pt} 0.054 &1.000 &  $-$0.10 & 0.11\\
 97219  & 0.828 &\hspace{4pt} 0.024 &1.000 &  $-$0.04 & 0.09\\
 97337  & 1.257 &\hspace{4pt} 0.131 &  -- &  $-$0.15 & 0.15\\
 97657  & 1.065 &$-$0.012 &1.000 &$\ \ $0.06 & 0.12\\
 97974  & 0.625 &\hspace{4pt} 0.163 &1.000 &  $-$0.24 & 0.21\\
 98655  & 0.766 &\hspace{4pt} 0.037 &1.000 &  $-$0.28 & 0.21\\
\noalign{\smallskip}\hline\noalign{\smallskip} 
\multicolumn{6}{c}{The remaining program stars}\\
\noalign{\smallskip}\hline\noalign{\smallskip}
 91128  & 1.394 &\hspace{4pt} 0.243&    -- &  $-$0.17 & 0.15\\
 92922  & 0.767 &\hspace{4pt} 0.012*&1.000 &$\ \ $0.19 & 0.16\\
 94145  & 0.242 &\hspace{4pt} 0.080*&1.200 &  $-$0.17 & 0.23\\
 94704  & 0.653 &\hspace{4pt} 0.289 &1.026 &  $-$1.52 & 0.52\\
 94743  & 0.282 &\hspace{4pt} 0.148 &1.208 &  $-$0.23 & 0.21\\
 94898  & 0.791 &$-$0.003* &1.000 &  $-$0.01 & 0.12\\
 95631  & 0.778 &$-$0.001* &1.000 &$\ \ $0.11 & 0.16\\
 95638  & 0.696 &\hspace{4pt} 0.063 &1.000 &  $-$0.07 & 0.15\\
 95843  & 0.418 &\hspace{4pt} 0.097 &1.161 &  $-$0.25 & 0.14\\
 96146  & 0.467 &$-$0.002* &1.000 &$\ \ $0.10 & 0.13\\
 97168  & 0.618 &\hspace{4pt} 0.140 &1.000 &  $-$0.40 & 0.13\\
 98381  & 1.080 &$-$0.024 &1.000 &  $-$0.02 & 0.06\\
 98829  & 0.712 &\hspace{4pt} 0.173 &1.053 &  $-$0.32 & 0.21\\
 99267  & 0.468 &\hspace{4pt} 0.323 &1.150 &  $-$1.62 & 0.13\\
\noalign{\smallskip}\hline 
\end{tabular}\\[1ex]
\parbox{3in}
{* Corrected for the luminosity effect as described in the text.} 
}                               
\end{center}                    
\end{table}

As can be seen from Fig.\ 3, all group 1 stars (including the subdwarf HIP\,99267, not 
plotted in the upper panel) lie either very close to the [Fe/H]--$\delta(0.6)$ lines 
(HIP\,96146, HIP\,93011, HIP\,95637, in order of decreasing [Fe/H]) or on the right-hand 
side of it (HIP\,94145, HIP\,95843, HIP\,94743, HIP\,99267, in order of increasing 
distance from the [Fe/H]--$\delta(0.6)$ lines). De-reddening the $B-V$ and $U-B$ 
color-indices would move a group 1 star to the right in Fig.\ 3. As an example, let us 
consider  HIP\,96146, the star represented in the upper panel of Fig.\ 3 by the leftmost 
open circle lying about half the standard error (equal to 0.016 mag) to the left of the 
[Fe/H]--$\delta(0.6)$ lines. If we assumed $E(B-V) = 0.05$ mag, the $\delta(0.6)$ would 
change from $-0.002$ mag to 0.033 mag. The open circle would then move to the right by 
two standard errors, so that the distance from the [Fe/H]--$\delta(0.6)$ lines would 
increase to 1.5 standard error. For the remaining group 1 stars, de-reddening would spoil 
the agreement with the [Fe/H]--$\delta(0.6)$ lines even more. We conclude that for group 
1 stars, including the two most distant program stars  HIP\,93011 ($r = 242\,\pm\,42$ pc) 
and HIP\,94145 ($r=229\,\pm\,36$ pc), their position in the $\delta(0.6)$, [Fe/H] plane 
implies $E(B-V) < 0.025$ mag. 

De-reddening a group 2 star moves it to the left in Fig.\ 3. Thus, de-reddening may move 
a group 2 star closer to the [Fe/H]--$\delta(0.6)$ lines if it lies on the right-hand 
side of them. Most such stars lie less than one standard error (of [Fe/H], $\delta(0.6)$, 
or both) off the lines. However, HIP\,95733 and HIP\,98829, represented by the two 
rightmost points in the upper panel of Fig.\ 3, deviate by more than three standard 
errors from the [Fe/H]--$\delta(0.6)$ lines. Shifting these two points so that they would 
lie between the two [Fe/H]--$\delta(0.6)$ lines would require $E(B-V) = 0.14$ mag for 
HIP\,95733 and $E(B-V) = 0.21$ mag for HIP\,98829. De-reddening HIP\,95733 and HIP\,98829 
using these numbers would make their $B-V$ color-indices much too blue for the stars' MK 
type, G8\,V in both cases (see Table 9 of Paper I). In addition, the two stars lie very 
close to the Lang's (1992) intrinsic relation in Fig.\ 1. We conclude that reddening 
cannot explain the rightward deviation of HIP\,95733 and HIP\,98829 from the 
[Fe/H]--$\delta(0.6)$ lines.

In addition to these two stars, and the group 1 star HIP\,94743 = V\,2077 Cyg (the open 
circle at $\delta(0.6) \approx 0.18$ in the upper panel of Fig.\ 3), the subdwarf, 
HIP\,99267 (the rightmost open square in the lower panel of Fig.\ 3) also deviates from 
the mean [Fe/H]--$\delta(0.6)$ relations by several standard errors in the abscissa. 
These deviations require explaining.

\vspace{0.5cm}
{\em 3.4. The $\beta$, $b-y$ Diagram}

\vspace{0.5cm}
In Fig.\ 4, the program stars for which the $\beta$ index has been measured by us are 
plotted in the $\beta$, $b-y$ plane using the values from Table 2, with the duplicity 
corrections for HIP\,94335, HIP\,94734 and HIP\,94743 (see Table 3) taken into account. 
The distances, $r$ (in parsecs) used in coding the symbols, were obtained from the 
revised Hipparcos parallaxes (van Leeuwen 2007). The solid line is the standard relation 
of Crawford (1975a) (with slight corrections due to Olsen 1988), valid for 
unreddened F0--G2 stars of luminosity classes III--V. The long-dashed line is Crawford's 
(1979) standard relation for A7--F0 stars. The short-dashed lines represent the 
envelopes of scatter in Crawford's (1975a) F-star $\beta$, $b-y$ diagram (his Fig.\ 
1); they show by how much an unreddened F-type star may deviate from the standard 
relation because of -- to use Crawford's term -- ``cosmic scatter.'' The arrow is the 
reddening vector $E(b-y) = 0.074$ mag, corresponding to $E(B-V) = 0.1$ mag (Crawford 
1975b). 

\begin{figure} 
\epsfbox{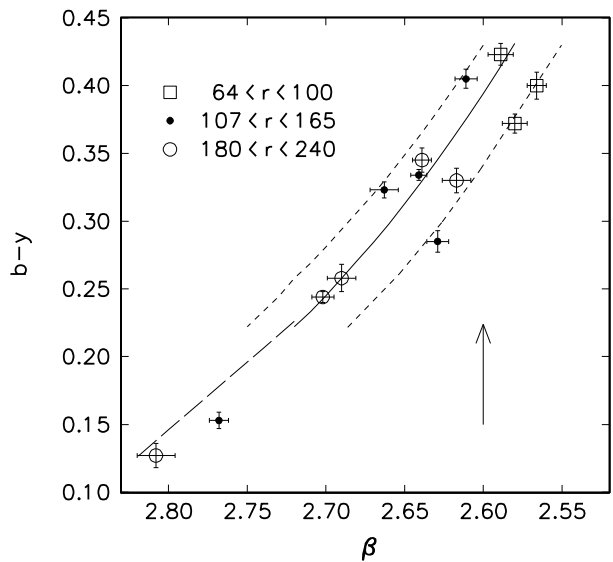} 
\caption{Program stars in the $\beta$, $b-y$ plane. The distances, $r$ (in parsecs) 
used in coding the symbols, were obtained from the revised Hipparcos parallaxes (van 
Leeuwen 2007). The solid line represents the standard relation of Crawford (1975a) (with 
slight corrections due to Olsen 1988), valid for unreddened F0--G2 stars of 
luminosity classes III--V. The long-dashed line is Crawford's (1979) standard relation 
for A7--F0 stars. The short-dashed lines show by how much an unreddened F-type star may 
deviate from the standard relation because of ``cosmic scatter'' (see Crawford 1975a, 
Fig.\ 1). The arrow is the reddening vector $E(b-y) = 0.074$ mag, corresponding to 
$E(B-V) = 0.1$ mag (Crawford 1975b).} 
\end{figure}

In Fig.\ 4, the deviations in the ordinate, $(b-y) - (b-y)_0$, where $(b-y)_0$ is the 
standard $b-y$ at the star's $\beta$, range from $-0.066$ mag for HIP\,99267, the F2 
sub- dwarf, to $0.026$ mag for HIP\,96146. The mean deviation is equal to 
$-0.015\,\pm\,0.009$ mag. However, HIP\,97168 (the rightmost open square in the figure), 
which shows the negative deviation of $-0.063$ mag, may do so because of (slight) 
emission in the H$\beta$ line at the time of observation (see the next subsection). If 
we rejected HIP\,97168 and the subdwarf, the mean deviation would become equal to 
$-0.006\,\pm\,0.008$ mag. For the five stars more distant than 180 pc, the mean 
deviation amounts to $-0.008\,\pm\,0.008$ mag. Since de-reddening would decrease the 
$b-y$\, deviations, we conclude that there is no evidence for reddening in Fig.\ 4. 

In Sect.\ 3.1 we noted that HIP\,94743, classified to F2 in Paper I, showed a deviation 
from the intrinsic relation in Fig.\ 1 which suggested a spectral type of about A9. In 
Fig.\ 4, this star (the leftmost filled circle) falls close to Crawford's (1979) 
standard relation for A7--F0 stars (the long-dashed line). 

\vspace{0.5cm}
{\em 3.5. The\, {\rm [Fe/H]}, $\delta m_1$ Diagrams}

\vspace{0.5cm} 
Crawford (1975a) has defined metal-abundance parameters $\delta m_1(\beta)$ 
and\linebreak \hbox{$\delta m_1(b-y)$}. In both cases, the parameter is a difference 
between the standard and observed value of $m_1$ (in the sense ''standard {\em minus\/} 
observed''); in case of $\delta m_1(\beta)$, the standard value is read from the 
standard relations between the $uvby \beta$ indices using the observed $\beta$ as the 
independent variable; in case of $\delta m_1(b-y)$, the independent variable is the 
observed $b-y$, corrected for reddening if necessary. 

Figs.\ 5 and 6 show the program stars in the [Fe/H], $\delta m_1(\beta)$ and the [Fe/H], 
$\delta m_1(b-y)$ planes using the [Fe/H] values from Table 9 of Paper I and the $uvby 
\beta$ indices from Table 2. For HIP\,94335, HIP\,94734 and HIP\,94743 the duplicity 
corrections (see Table 3) were taken into account. Zero reddening was assumed in drawing 
both figures. Fig.\ 6 shows one more program star than Fig.\ 5 does, viz., HIP\,94704, 
the G0 subdwarf for which $\beta$ was not measured (the open square with the long 
horizontal error bar). For most stars, we used the F-star standard $m_1-\beta$ and 
$m_1-(b-y)$ relations from Table I of Crawford (1975a); for HIP\,97168 
(the rightmost open square in Fig.\ 4) the standard $m_1-\beta$ relation had to be 
extrapolated. For two stars, HIP\,94743 and HIP\,94145 (the leftmost filled circle and 
the leftmost open circle in Fig.\ 4, respectively) we used Crawford's (1979) A-star 
standard relation. 
\begin{figure} 
\epsfbox{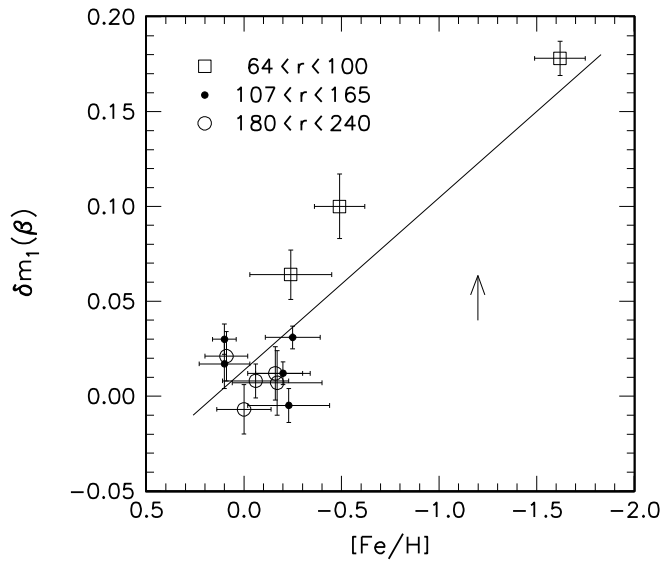} 
\caption{Program stars in the [Fe/H], $\delta m_1(\beta)$ plane. The symbols are coded 
as in Fig.\ 4. The line represents the [Fe/H]--$\delta m_1(\beta)$ relation of 
Crawford \& Perry (1976). The arrow is the reddening vector $E(m_1) = -0.024$ mag, 
corresponding to $E(B-V) = 0.1$ mag (Crawford 1975b).} 
\end{figure}

In Figs.\ 5 and 6 all points in common but one show similar deviations from the 
[Fe/H]--$\delta m_1$ relations of Crawford \& Perry (1976). The exception is HIP\,97168 
(the open square at [Fe/H] $= -0.5$) for which $\delta m_1(\beta) = 0.100$ mag while 
$\delta m_1(b-y) = 0.024$ mag. The discrepancy would disappear if the star's $\beta$ 
were increased by 0.030 mag. This is suggests a slight emission at H$\beta$ at the time 
of observation.

\begin{figure} 
\epsfbox{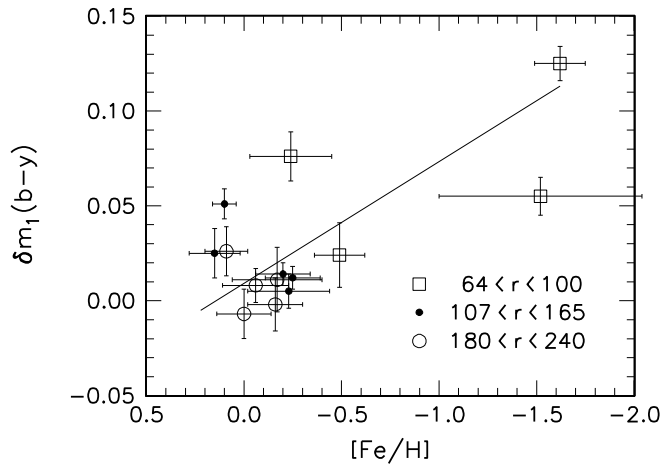} 
\caption{Program stars in the [Fe/H], $\delta m_1(b-y)$ plane. The symbols are coded as 
in Fig.\ 4. The line represents the [Fe/H]--$\delta m_1(b-y)$ relation of Crawford \& 
Perry (1976).} 
\end{figure}

\vspace{0.5cm}
\centerline{\bf 4. The Effective Temperatures}

\vspace{0.5cm} 
{\em 4.1. Photometric Effective Temperatures}

\vspace{0.5cm}
For all Population I program stars we can obtain $T_{\rm eff}$ from the $B-V$ 
color-index. In case of the stars for which we have measured the $\beta$ index, we can 
also derive $T_{\rm eff}$ from this index. We shall denote these photometric effective 
temperatures by $T_{\rm eff}(B-V)$ and $T_{\rm eff}(\beta)$, respectively. In case of the  
subdwarfs, $T_{\rm eff}$ is also a function of a metallicity parameter (see below).

For deriving $T_{\rm eff}(B-V)$, we used the $B-V$ color-indices from Table 1 (plus the 
duplicity corrections in case of the binaries HIP\,94335, HIP\,94734 and HIP\,94743) 
and Flower's (1996) $T_{\rm eff}$ vs.\ $(B-V)$ calibration (his Table 3). For $T_{\rm 
eff}(\beta)$, we used the $uvby \beta$ indices from Table 2 (with the duplicity 
corrections in case of the binaries) and the FORTRAN program UVBYBETA\footnote{Written 
in 1985 by T.T.\ Moon of the University London and modified in 1992 by R.\ Napiwotzki of 
Universitaet Kiel. The program is based on the grid published in Moon \& Dworetsky 
(1985)}, kindly made available to us by Dr.\ Napiwotzki. In both cases, we assumed the 
color indices to be unreddened. The results are given in Table 5.

\begin{table} 
\begin{center}
\centerline{T a b l e \quad 5}
\centerline{The photometric effective temperatures}
\vspace{0.3cm}
{\small
\begin{tabular}         
{ccrcr} \\
\hline\noalign{\smallskip}
HIP    & $T_{\rm eff}(B-V)$ & s.e. & $T_{\rm eff}(\beta)$ & s.e.\\
\noalign{\smallskip}\hline\noalign{\smallskip}
\multicolumn{5}{c}{PATS}\\
\noalign{\smallskip}\hline\noalign{\smallskip}
 93011  &  6720 & 18& 6810   &  93  \\
 94335  &  6232 & 17& 6423    &  55  \\
 94497  &  4980 &  6& --  &  --  \\
 94565  &  6175 & 16& 6340   &  103  \\
 94734  &  5856 & 14& 6029   &   92  \\
 95098  &  6130 & 16& 6105   &  116  \\
 95637  &  6853 & 14& 6925   &   37  \\
 95733  &  5411 & 14& --  &  --  \\
 96634  &  5267 & 10& --  &  --  \\
 96735  &  5114 &  9& --  &  --  \\
 97219  &  5212 & 10& --  &  --  \\
 97337  &  4387 &  7& --  &  --  \\
 97657  &  4722 &  9& --  &  --  \\
 97974  &  5802 & 14& 5810   &  115  \\
 98655  &  5370 & 13& --  &  --  \\
\noalign{\smallskip}\hline\noalign{\smallskip} 
\multicolumn{5}{c}{The remaining program stars}\\
\noalign{\smallskip}\hline\noalign{\smallskip}
 91128  &  4163 &  8& --  &  --  \\
 92922  &  5367 & 11& --  &  --  \\
 94145  &  7525 & 11& 7754    &  82  \\
 94704  &  5182 & 11& 5490  &  48  \\
 94743  &  7547 &133& 7449    &  56  \\
 94898  &  5305 &  5& --  &  --  \\
 95631  &  5339 & 10& --  &  --  \\
 95638  &  5571 &  9& --  &  --  \\
 95843  &  6642 &  9& 6229   &  85  \\
 96146  &  6422 & 22& 6576  &   9  \\
 97168  &  5827 & 14& --  &  --  \\
 98381  &  4695 &  7& --  &  --  \\
 98829  &  5523 & 17& --  &  --  \\
 99267  &  5639 & 28& 5748  &  41  \\
\noalign{\smallskip}\hline 
\end{tabular}\\[1ex]
\parbox{2.8in}
{For HIP\,94704 and HIP\,99267 $T_{\rm eff}(B-V,\delta(0.6))$ is given instead of $T_{\rm 
eff}(B-V)$, and $T_{\rm eff}(b-y,{\rm [Fe/H]})$, instead of $T_{\rm eff}(\beta)$.} 

}
\end{center}                     
\end{table}

The program UVBYBETA requires all four $uvby \beta$\, indices as input. However, for F 
and G stars, the effective temperature is derived from the $\beta$\, index, with a 
marginal contribution from $c_1$. Therefore, $T_{\rm eff}(\beta)$ was not obtained for 
HIP\,97168 because of the suspected emission at H$\beta$ (see Sect.\ 3.5).  

For the two subdwarfs, HIP\,94704 and HIP\,99267, we computed $T_{\rm eff}$ from $B-V$ 
and $\delta(0.6)$ using the calibration of Carney et al.\ (1994), and from $b-y$ and 
[Fe/H] using the calibration of Spite et al.\ (1996). The color indices were assumed to 
be unreddened. These effective temperatures we shall refer to as $T_{\rm 
eff}(B-V,\delta(0.6))$ and $T_{\rm eff}(b-y,{\rm [Fe/H]})$, respectively. They are also 
listed in Table 5. 

\begin{figure} 
\epsfbox{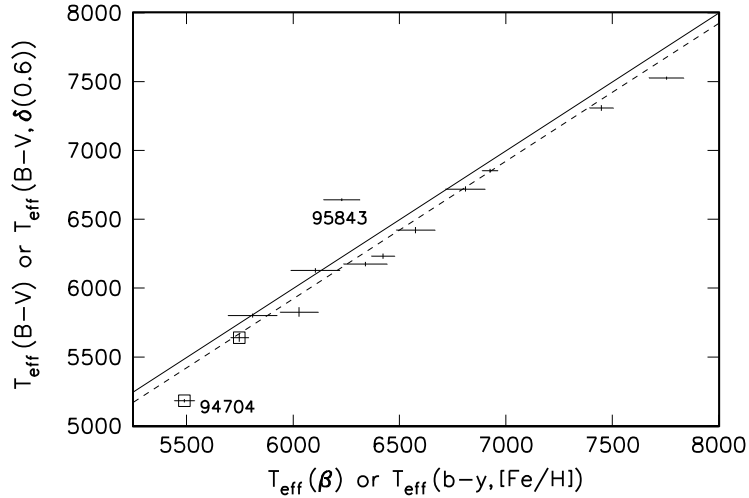} 
\caption{Comparison of the photometric effective temperatures. Population I stars are 
shown with the crossed error bars, and the subdwarfs, with open squares. The solid line 
has unit slope and zero intercept. The short-dashed line corresponds to the straight 
mean difference $T_{\rm eff}(B-V) - T_{\rm eff}(\beta)$. HIP numbers label points that 
deviate from the short-dashed line by more than $3\sigma$, where $\sigma$ is the standard 
deviation of an individual temperature difference (very nearly a half of the 
corresponding horizontal error bar).} 
\end{figure}

A comparison of the photometric effective temperatures is shown in Fig.\ 7, where $T_{\rm 
eff}(B-V)$ is plotted as a function of $T_{\rm eff}(\beta)$ for Population I stars, and 
$T_{\rm eff}(B-V,\delta(0.6))$ is plotted as a function of $T_{\rm eff}(b-y,{\rm 
[Fe/H]})$ for the subdwarfs. The straight mean difference $T_{\rm eff}(B-V) - T_{\rm 
eff}(\beta)$ is equal to $-74\,\pm\,54$ K. In the figure, the short-dashed line is 
shifted by this number from the solid line which has unit slope and zero intercept. The 
points that deviate from the short-dashed line by more than $3\sigma$, where $\sigma$ is 
the standard deviation of an individual temperature difference (very nearly a half of 
the corresponding horizontal error bar) are labeled with the HIP numbers. 

\vspace{0.5cm} 
{\em 4.2. Comparison with Spectroscopic Effective Temperatures}

\vspace{0.5cm} 
In Paper I, we have derived the spectroscopic effective temperatures of program stars by 
two methods. In the present discussion we shall call these methods (1) and (2). In method 
(1), we compared the spectrograms, obtained with the Catania Astrophysical Observatory's 
fiber-fed echelle spectrograph FRESCO on a 91-cm telescope, with spectrograms of 
reference stars with known atmospheric parameters. In addition to $T_{\rm eff}$, this 
method yielded $\log g$, [Fe/H] and the MK type. We used two independent grids of 
reference stars. One was based on spectrograms selected from an archive known as ELODIE 
(Prugniel \& Soubiran 2001), and the other, on reference stars observed with FRESCO. We 
shall refer to the effective temperatures derived with the ELODIE grid as $T_{\rm 
eff}(1{\rm E})$, and those derived with the FRESCO grid, by $T_{\rm eff}(1{\rm F})$. In 
Paper I, $T_{\rm eff}(1{\rm E})$ and the remaining atmospheric parameters obtained with 
the ELODIE grid are listed in Table 9, and those obtained with the FRESCO grid, in Table 
10. Table 9 contains all program star, while Table 10, only 16, having $T_{\rm eff}$ in 
the range from about 5000 K to about 6000 K. The MK types and the metallicity parameters 
[Fe/H] already discussed in the present paper have been taken from Table 9. 

In method (2), we used echelle spectrograms obtained with the CfA Digital Speedometers at 
the Oak Ridge Observatory, Harvard, Massachusetts and the F.L.\ Whipple Observatory, 
Mount Hopkins, Arizona. From these spectrograms, $T_{\rm eff}$ and $\log g$ were 
determined by cross-correlation with templates computed from model stellar atmospheres. 
We shall refer to the effective temperatures derived by this method as $T_{\rm eff}(2)$. 
In Paper I, $T_{\rm eff}(2)$ are listed in Table 11. 

The spectroscopic effective temperatures $T_{\rm eff}(1{\rm E})$ are plotted in Fig.\ 8 
vs.\linebreak $T_{\rm eff}(B-V)$ for Population I stars, and vs.\ $T_{\rm 
eff}(B-V,\delta(0.6))$ for the subdwarfs. As can be seen from the figure, at $T_{\rm 
eff}(B-V) < 6250$ K the points scatter randomly around the line of unit slope and zero 
intercept, with only the subdwarfs showing large deviations. For $T_{\rm eff}(B-V) < 
6250$ K, the mean difference $T_{\rm eff}(1{\rm E}) - T_{\rm eff}(B-V)$ amounts to an 
insignificant $-15\,\pm\,17$ K. However, for $T_{\rm eff}(B-V) > 6400$ K all points fall 
below the line. For these points, the mean difference $T_{\rm eff}(1{\rm E}) - T_{\rm 
eff}(B-V)$ is equal to $-311\,\pm\,34$ K; this mean difference was used to draw the 
short-dashed line in Fig.\ 8. If $T_{\rm eff}(\beta)$ were used instead of $T_{\rm 
eff}(B-V)$, the mean difference for $T_{\rm eff}(B-V) > 6400$ K would be equal to 
$-356\,\pm\,90$ K. Since $T_{\rm eff}(\beta)$ and $T_{\rm eff}(B-V)$ were obtained 
independently of each other, the systematic deviation for $T_{\rm eff}(B-V) > 6400$ K 
must be due to a problem in deriving $T_{\rm eff}(1{\rm E})$ for the six hottest stars 
in our sample. 

\begin{figure} 
\epsfbox{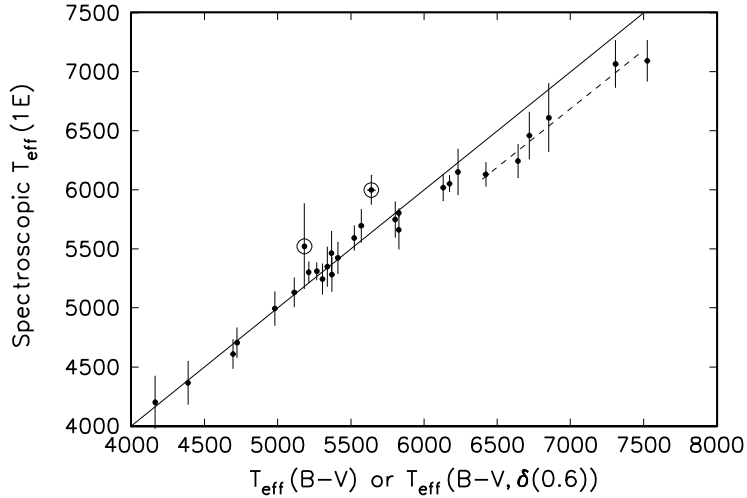} 
\caption{The spectroscopic effective temperature $T_{\rm eff}(1{\rm E})$ plotted as a function 
of $T_{\rm eff}(B-V)$ for Population I stars (points), and $T_{\rm eff}(B-V,\delta(0.6))$ 
for the subdwarfs (encircled points). HIP\,94743 (the rightmost point) was shifted up 
and to the right to avoid coincidence with HIP\,94145. The 
horizontal error bars are too short to show up. The solid line has unit slope and zero 
intercept; the short-dashed line runs 311 K below it.} 
\end{figure}

The effective temperatures obtained with the FRESCO grid may suffer from the same 
problem because for the two stars with $T_{\rm eff}(B-V) > 6400$ K, HIP\,95843 and 
HIP\,96146, for which $T_{\rm eff}(1{\rm F})$ were derived, the mean difference $T_{\rm 
eff}(1{\rm F})\ -$ $T_{\rm eff}(B-V)$ is equal to $-370$ K. 

A comparison of the spectroscopic effective temperatures based on model stellar 
atmospheres, $T_{\rm eff}(2)$, with the photometric effective temperatures is shown in 
Fig.\ 9. In this figure, most points lie close to the line of unit slope and zero 
intercept, although the scatter is larger than in Fig.\ 8. The mean difference  for 
$T_{\rm eff}(B-V) > 6400$ K is equal to $-112\,\pm\,64$ K. This value differs from zero 
by less than 2$\sigma$ suggesting that $T_{\rm eff}(2)$ does not suffer from the problem 
of underestimating effective temperatures of the hottest stars. The overall mean 
difference $T_{\rm eff}(2) - T_{\rm eff}(B-V)$ is equal to $-82\,\pm\,35$ K. Thus, the 
$T_{\rm eff}(2)$ scale is slightly cooler from the $T_{\rm eff}(B-V)$ scale. 

\begin{figure} 
\epsfbox{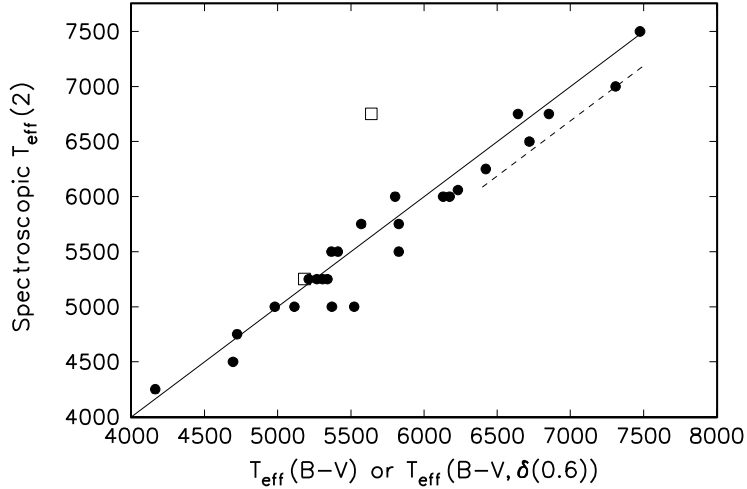} 
\caption{The spectroscopic effective temperature $T_{\rm eff}(2)$ plotted as a function 
of $T_{\rm eff}(B-V)$ for Population I stars (circles), and $T_{\rm 
eff}(B-V,\delta(0.6))$ for the subdwarfs (open squares). The lines are from Fig.\ 8. The 
discrepant open square is HIP\,99267.} 
\end{figure}

\vspace{0.5cm}
\centerline{\bf 5. The Photometric Surface Gravities}

\vspace{0.5cm} 
Using the $uvby \beta$ indices from Table 2 (with the duplicity corrections for the 
binaries taken into account) and the program UVBYBETA, we have obtained the photometric 
surface gravities listed in Table 6. We have denoted these values by $\log g(c_1,\beta)$ 
to indicate their sensitivity to $c_1$ and $\beta$. No reddening corrections were 
applied. Because of the sensitivity to $\beta$, HIP\,97168 was omitted; HIP\,99267 was 
omitted because the program is not calibrated for subdwarfs. 

\begin{table} 
\begin{center}
\centerline{T a b l e \quad 6}
\centerline{The photometric surface gravities}
\vspace{0.3cm}
{\small
\begin{tabular}         
{ccr} \\
\hline\noalign{\smallskip}
HIP    & $\log g(c_1,\beta)$ & s.e.\\
\noalign{\smallskip}\hline\noalign{\smallskip}
\multicolumn{3}{c}{PATS}\\
\noalign{\smallskip}\hline\noalign{\smallskip}
 93011  &  4.27 & 0.10  \\
 94335  &  4.62 & 0.06  \\
 94565  &  4.02 & 0.16  \\
 94734  &  3.89 & 0.11  \\
 95098  &  3.96 & 0.20  \\
 95637  &  4.41 & 0.04  \\
 97974  &  4.74 & 0.07  \\
\noalign{\smallskip}\hline\noalign{\smallskip} 
\multicolumn{3}{c}{The remaining program stars}\\
\noalign{\smallskip}\hline\noalign{\smallskip}
 94145  &  4.16 & 0.06  \\
 94743  &  4.45 & 0.05  \\
 95843  &  3.94 & 0.12  \\
 96146  &  4.20 & 0.13  \\
\noalign{\smallskip}\hline 
\end{tabular}

}
\end{center}                     
\end{table}

\begin{figure} 
\epsfbox{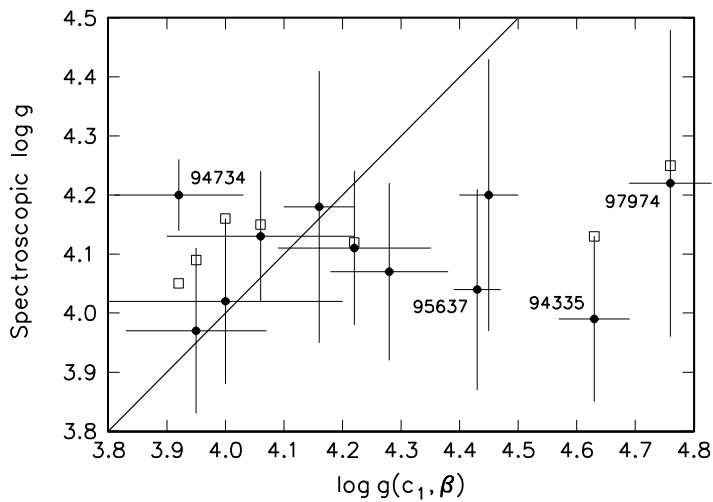} 
\caption{The spectroscopic $\log g$ plotted as a function of the photometric $\log 
g(c_1, \beta)$. Filled circles have the ordinates obtained with the ELODIE grid, and 
open squares, those obtained with the FRESCO grid. The solid line has unit slope and 
zero intercept. Deviant points are labeled with HIP numbers.} 
\end{figure}

A comparison of the spectroscopic $\log g$ with $\log g(c_1,\beta)$ is shown in Fig.\ 
10. Filled circles have the ordinates obtained with the ELODIE grid (see Paper I, Table 
9), and open squares, those obtained with the FRESCO grid (Paper I, Table 10). Deviant 
points are labeled with HIP numbers. The overall agreement between the spectroscopic 
$\log g$ and $\log g(c_1,\beta)$ is unsatisfactory in both cases. 

\vspace{0.5cm}
\centerline{\bf 6. Summary}

\vspace{0.5cm} 
We find no evidence that the program stars are reddened. In this we disagree with KIC-10 
which gives $E(B-V)$ ranging from $0.01$ to $0.06$ mag for nine of our program stars. 
Unfortunately, a detailed star-by-star comparison must be postponed until the catalog 
is made public. 

The photometric effective temperatures derived in Sect.\ 4.1 from $B-V$,\linebreak 
\hbox{$T_{\rm eff}(B-V)$}, agree very well with the spectroscopic effective 
temperatures given in Table 9 of Paper I, for $T_{\rm eff}(B-V) < 6250$ K. For $T_{\rm 
eff}(B-V) > 6400$ K, these spectroscopic effective temperatures, referred to in the 
present paper as $T_{\rm eff}(1{\rm E})$, are systematically smaller from the $T_{\rm 
eff}(B-V)$ by $311\,\pm\,34$ K (see Fig.\ 8). The photometric effective temperatures 
$T_{\rm eff}(\beta)$, derived in Sect.\ 4.1 from $\beta$, show very nearly the same 
problem: for $T_{\rm eff}(B-V) > 6400$ K, the difference $T_{\rm eff}(1{\rm E}) - T_{\rm 
eff}(\beta)$  amounts to $-346\,\pm\,91$ K. Clearly, the calibrations of the $T_{\rm 
eff}(1{\rm E})$, $T_{\rm eff}(B-V)$ and $T_{\rm eff}(\beta)$ scales for F-type stars 
should be examined. We leave this as a task for the (near) future. 

The photometric surface gravities of the program stars, $\log g(c_1,\beta)$, derived in 
Sect.\ 5 from $c_1$ and $\beta$, range from $3.89\,\pm\,0.11$ to $4.74\,\pm\,0.07$. The 
range of the spectroscopic $\log g$ of Paper I is a factor of two smaller (see Fig.\ 
10). Whether this is caused by incorrect spectroscopic $\log g$ values of HIP\,94734, 
HIP\,94335 and HIP\,97974 or by a an error in the photometric surface gravities needs 
looking into. 

\vspace{0.5cm}
\centerline{APPENDIX}

\vspace{0.5cm} 
\centerline{\bf Photometric Reductions}

\vspace{0.5cm} 
For the $UVB$ reductions we have used the following equations:
\begin{equation}
A_1 + B_1 (B-V) + K_1 X = C_{\rm BV}^{\rm X} + 0.03 (B-V) X,
\end{equation}
\begin{equation}
A_2 + B_2 (U-B) + K_2 X = C_{\rm UB}^{\rm X},
\end{equation}
and
\begin{equation}
A_3 + B_3 (B-V) + K_3 X = V^{\rm X} - V,
\end{equation}
where X is the air mass, $C_{\rm BV}^{\rm X}$ is the raw blue-visual color index, 
$C_{\rm UB}^{\rm X}$ is the raw ultraviolet-blue color index, $V^{\rm X}$ is the raw 
visual magnitude, $A_1$, $A_2$, $A_3$ and $B_1$, $B_2$, $B_3$ are the transformation 
coefficients, and $K_1$, $K_2$, $K_3$ are the atmospheric extinction coefficients. In 
writing these equations we have assumed (1) the second-order extinction coefficient for 
the blue-visual color index to be equal to $-0.03$, (2)~the second-order extinction 
coefficient for the ultraviolet-blue color index to be equal to zero, and (3) the 
second-order extinction coefficient for the visual magnitude to be equal to zero. 

Each $UBV$ observation of a standard star, consisting of $X$, $C_{\rm BV}^{\rm X}$, 
$C_{\rm UB}^{\rm X}$ and $V^{\rm X}$, was used to form three independent equations of 
condition, corresponding to Eqs.\ (1), (2) and (3). For a number of standard stars 
observed at a time, the equations of conditions were solved by the method of least 
squares using the standard color-indices and the standard magnitudes listed in columns 
3, 4 and 2 of Table 7. Thus, Eqs.\ (1) yielded $A_1$, $B_1$ and $K_1$. These 
coefficients were then used to compute the $B-V$ color-indices of program stars from 
$C_{\rm BV}^{\rm X}$ and $X$; because of the second term on the r.h.s of Eq.\ (1), this 
was done by iteration. In the next step, $A_3$, $B_3$ and $K_3$, which resulted from 
solving Eqs.\ (3), and the $B-V$ just obtained were used to compute the $V$ magnitudes 
of the program stars from $V^{\rm X}$ and $X$. The $U-B$ indices of the program stars 
were computed from $C_{\rm UB}^{\rm X}$ and $X$ using $A_2$, $B_2$ and $K_2$. On most 
nights, observations of the standard stars were made before and after program stars' 
observations, so that the coefficients could be interpolated in order to compensate for 
their variation during the night. 

In case of one standard star, HD\,157881, the deviations from the solutions of Eqs.\ (1) 
were large but consistent, their mean value amounting to $0.046$ mag. We concluded that 
the $B-V$ index of this star given in reference (1) must be in error and adjusted it 
accordingly; the adjusted value is indicated in Table 7 with a colon. 

In case of the $uvby$ reductions, the equations for $b-y$, $m_1$ and $y$ had the same 
form as the $UBV$ equations just discussed, except that there was no need to include a 
second-order $b-y$ extinction coefficient. Thus, the equations were the following:
\begin{equation}
a_1 + b_1 (b-y) + k_1 X = C_{\rm by}^{\rm X},
\end{equation}
\begin{equation}
a_2 + b_2 m_1 + k_2 X = C_{\rm m}^{\rm X},
\end{equation}
and
\begin{equation}
a_3 + b_3 (b-y) + k_3 X = y^{\rm X} - V,
\end{equation}
where X is the air mass, $C_{\rm by}^{\rm X}$ is the raw $b-y$ color index, 
$C_{\rm m}^{\rm X}$ is the raw $m_1$ index, $y^{\rm X}$ is the raw $y$
magnitude, $a_1$, $a_2$, $a_3$ and $b_1$, $b_2$, $b_3$ are the transformation 
coefficients, and $k_1$, $k_2$, $k_3$ are the atmospheric extinction coefficients. 

\begin{table}[t]
\begin{center}
\centerline{T a b l e \quad 7}
{Adopted standard magnitudes and color indices}
\vspace{0.3cm}
{\small
\begin{tabular}
{clrrcrllcc} \\
\hline\noalign{\smallskip}
HD  &\ \ $V$ & $B-V$ & $U-B$ & Ref. & $b-y$ &\ \,$m_1$ &\ \,$c_1$ & $\beta$ & Ref.\\
\noalign{\smallskip}\hline\noalign{\smallskip}
  142860 & 3.845* &   0.478\hspace*{3pt}& $-$0.025 & (1) &    0.319 & 0.150   & 0.401\& & 2.633 & (3)\\
  144206 & 4.738  &$-$0.101\hspace*{3pt}& $-$0.321 & (1) & $-$0.032 & 0.105   & 0.756\& & 2.756 & (3)\\
  146470 & 8.430  &   1.350\hspace*{3pt}&    1.510 & (1) &    0.854 & 0.561\$ & 0.482   &  ---  & (3)\\
  149801 &\hspace*{5pt}---   &    ---\hspace*{7pt}&     ---\hspace*{5pt}&     & $-$0.014 & 0.141\$ & 0.902   &  ---  & (3)\\
  151288 & 8.102  &   1.369\hspace*{3pt}&    1.289 & (1) &    0.784 & 0.757\$ & 0.030   & 2.508 & (3)\\
  154029 & 5.268  &   0.023\hspace*{3pt}&    0.027 & (1) &    0.001 & 0.172   & 1.102\& & 2.885 & (3)\\
  157214 & 5.393  &   0.619\hspace*{3pt}&    0.069 & (1) &    0.404 & 0.179   & 0.312\& & 2.590 & (3)\\
  157881 & 7.543* &   1.314:&1.260&(1)& ---\hspace*{5pt}&\hspace*{6pt}---&\hspace*{6pt}---&  ---  &    \\
  158148 & 5.525  &$-$0.135\hspace*{3pt}& $-$0.580 & (1) & $-$0.040 & 0.091   & 0.435\& & 2.688 & (3)\\
  160346 & 6.533* &   0.954\hspace*{3pt}&    0.778 & (1) &---\hspace*{5pt}&\hspace*{6pt}---    &\hspace*{6pt}---    &  ---  &    \\
  160365 & 6.116* &   0.556\hspace*{3pt}&     ---\hspace*{5pt}& (1) &    0.374 & 0.162   & 0.555\& & 2.634 & (3)\\
  164058 & 2.242* &   1.517\hspace*{3pt}&    1.882 & (2) &    0.941 & 0.811\$ & 0.373   &  ---  & (3)\\
  165401 & 6.806\#&    --- \hspace*{5pt}&     ---\hspace*{5pt}& (1) &    0.393 & 0.166   & 0.288\& & 2.580 & (3)\\
  176486 & 7.257\#&    --- \hspace*{5pt}&     ---\hspace*{5pt}& (3) &    1.125 & 0.735\$ & 0.310   &  ---  & (3)\\
  178233 & 5.531  &   0.290\hspace*{3pt}&    0.040 & (1) &    0.176 & 0.190   & 0.747\& & 2.758 & (3)\\
  184171 & 4.739  &$-$0.142\hspace*{3pt}& $-$0.658 & (1) & $-$0.057 & 0.095   & 0.376\& & 2.656 & (3)\\
  185395 & 4.475  &   0.382\hspace*{3pt}& $-$0.029 & (1) &    0.261 & 0.157   & 0.502\& & 2.688 & (3)\\
  186429 & 7.552\#&    --- \hspace*{5pt}&     ---\hspace*{5pt}& (3) &    0.855 & 0.607\$ & 0.496   &  ---  & (3)\\
  188665 & 5.140  &$-$0.136\hspace*{3pt}& $-$0.550 & (1) & $-$0.059 & 0.098   & 0.453\& & 2.715 & (3)\\
  190993 & 5.068  &$-$0.177\hspace*{3pt}& $-$0.692 & (1) & $-$0.076 & 0.100   & 0.296\& & 2.686 & (3)\\
  195943 & 5.381  &   0.079\hspace*{3pt}&    0.046 & (1) &    0.023 & 0.205   & 0.980\& & 2.920 & (3)\\
  196035 & 6.470  &$-$0.140\hspace*{3pt}& $-$0.670 & (1) & $-$0.060 & 0.087   & 0.287\& & 2.688 & (3)\\
  196090 & 7.787  &   1.412\hspace*{3pt}&    1.560 & (1) &    0.901 & 0.596\$ & 0.444   &  ---  & (4)\\
  198639 & 5.061  &   0.198\hspace*{3pt}&    0.123 & (1) &    0.108 & 0.208   & 0.897\& & 2.843 & (3)\\
  201891 & 7.371  &   0.511\hspace*{3pt}& $-$0.158 & (1) &    0.358 & 0.104   & 0.262\& & 2.590 & (3)\\
  202575 & 7.950  &   1.044\hspace*{3pt}&    0.840 & (1) &    0.581 & 0.568\$ & 0.205   &  ---  & (3)\\
  207978 & 5.532  &   0.413\hspace*{3pt}& $-$0.125 & (1) &    0.299 & 0.122   & 0.425\& & 2.640 & (3)\\
  216397 & 4.983  &   1.559\hspace*{3pt}&    1.926 & (1) &    1.000 & 0.750\$ & 0.450   &  ---  & (5)\\
\noalign{\smallskip}\hline                                                               
\end{tabular}\\
\parbox{5in}
{* Not used in the $y$ reductions.\\
: Adjusted (see text).\\
\# Used in the y reductions, but not in the $V$ reductions.\\
\$ Not used in the final $m_1$ reductions.\\ 
\& Used in solving Eqs. (7).\\
(1) Mermilliod (1991).\\ 
(2) Oja (1991).\\
(3) Hauck \& Mermilliod (1998).\\
(4) Olsen (1993).\\
(5) Olson (1974).}                                                                           
}                               
\end{center}
\end{table}

Eqs.\ (4), (5) and (6) were solved using the raw color-indices and the raw $y$ 
magnitudes of the standard stars and the standard values listed in Table 7. In case of 
$m_1$, HD\,149801 and most standard stars redder than $b-y = 0.5$ mag showed large 
deviations (of about $\pm 0.05$ mag) from the solution, while the remaining standard 
stars showed deviations smaller than about $\pm 0.01$ mag. Therefore, we solved Eqs.\ 
(5) again, using only the standard $m_1$ values of these remaining standard stars. In 
Table 7, the $m_1$ values not used in this final solution are indicated with a dollar 
sign. 

The $c_1$ reductions were began with solving the equations
\begin{equation}
a_4 + b_4 c_1 + k_4 X = C_1^{\rm X},
\end{equation}
where $C_1^{\rm X}$ is the raw $c_1$ index and the remaining symbols are analogous to 
those in the former equations, using all standard $c_1$ values listed in column 8 of 
Table 7. However, the solution had an excessively large standard deviation. This was 
caused by the same standard stars that did not fit the $m_1$ solution. Therefore, we 
rejected these standards and solved Eqs.\ (7) using standard $c_1$ values of the 
remaining ones, indicated with an ampersand in column 8 of Table 7. The transformation 
coefficients $a_4$ and $b_4$, and the extinction coefficient $k_4$ obtained in this way 
yielded the deviations plotted as a function of $b-y$ in Fig.\ 11.  

\begin{figure} 
\epsfbox{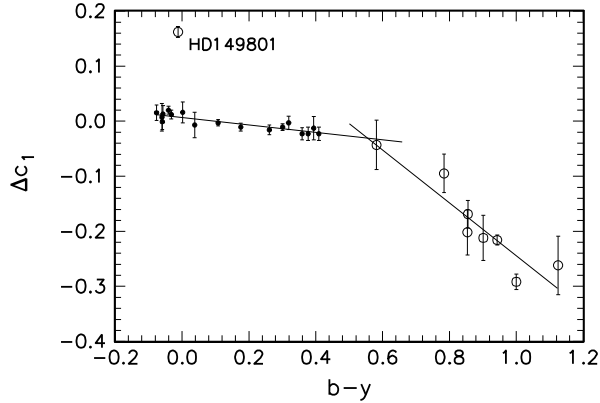} 
\caption{The $c_1$ deviations for the standard stars (in the sense ``computed {\em 
minus\/} standard'') plotted as a function of standard $b-y$. Filled circles are 
standard stars the $c_1$ indices of which were used in solving Eqs.\ (7), while open 
circles are the remaining standard stars. The straight lines were fitted to the points by 
the method of least squares.} 
\end{figure}

As can be seen from Fig.\ 11, the deviations can be represented by two straight lines, a 
less inclined one for $b-y < 0.6$ mag, and a more inclined one to the red of $b-y = 0.6$ 
mag. Taking this into account, we reduced the $c_1$ observations of program stars in two 
steps. First, we computed intermediate values of $c_1$ using Eq.\ (7) and the extinction 
and transformation coefficients obtained from the solution with the limited number of 
standards, and then, we corrected these intermediate $c_1$ indices using the slopes and 
zero-intersect values of the lines shown in Fig.\ 11. 

Finally, in the $\beta$ reductions we have used the eqution
\begin{equation}
a_5 + b_5 \beta  = \beta',
\end{equation}
where $\beta$ are the standard values from column 9 of Table 7 and $\beta'$ are the raw 
values. 

\vspace{0.5cm}{\bf Acknowledgments.} 
This work was partly supported by MNiSW grant N203 014 31/2650 and the University of 
Wroc{\l}aw grant N\b{o} 2646/W/IA/06. JM-\.Z thanks the Danish Natural Science Research 
Council, the Italian National Institute for Astrophysics (INAF), and the University of 
Catania for financial support.

\vspace{0.5cm}
\centerline{REFERENCES}
\vspace{0.3cm}
{\small

{Boesgaard, A.M., and Friel, E.D.} {1990} {Astrophys.\ J.} {351} {467}

{Cameron, L.M.} {1985} {Astron.\ Astrophys.} {146} {59}

{Carney, B.W., Latham, D.W., Laird, J.B., and Aguilar, L.A.} {1994} {Astron.\ J.} {107} 
{2240} 

{Crawford, D.L.} {1975a} {Astron.\ J.} {80} {955}

{Crawford, D.L.} {1975b} {PASP} {87} {481}

{Crawford, D.L.} {1979} {Astron.\ J.} {84} {1858}

{Crawford, D.L., and Perry, C.L.} {1976} {PASP} {88} {454}

{ESA} {1997} {``The Hipparcos and Tycho Catalogues''} {ESA SP-1200}

{Flower, P.J.} {1996} {Astrophys.\ J.} {469} {355}

{Hauck, B., and Mermilliod, M.} {1998} {Astron.\ Astrophys.\ Suppl.\ Ser.} {129} {431}

{Johnson, H.L.} {1966} {Ann.\ Rev.\ Astron.\ Astrophys.} {4} {193}

{Johnson, H.L., and Knuckles, C.F} {1955} {Astrophys.\ J.} {122} {209}

{Karata\c{s}, Y., and Schuster, W.J.} {2006} {MNRAS} {371} {1793}

{Lang, K.R.} {1992} {Astrophysical Data} {Springer} 

{Mendoza, E.E.} {1967} {Bol.\ Obs.\ Tonantzintla Tacubaya} {4} {149}

{Mermilliod, J.-C.} {1991} {Catalogue of Homogeneous Means in the UBV System} 
{Institut d'Astronomie, Universite de Lausanne}

{Molenda-\.Zakowicz, J., Frasca, A., Latham, D.W., and Jerzykiewicz, M.} {2007} 
{Acta Astron.} {57} {301} {Paper I}

{Moon, T.T, and Dworetsky, M.M} {1985} {MNRAS} {217} {305}

{Oja, T.} {1991} {Astron.\ Astrophys.\ Suppl.\ Ser.} {89} {415}

{Olsen, E.H.} {1988} {Astron.\ Astrophys.} {189} {173}

{Olsen, E.H.} {1993} {Astron.\ Astrophys.\ Suppl.\ Ser.} {102} {89}

{Olson, E.C.} {1974} {Astron.\ J.} {79} {1424}

{Pinsonneault, M.H., Terndrup, D.M., Hanson, R.B., and Stauffer, J.R.} {2004} 
{Astrophys.\ J.} {600} {946}

{Popper, D.M., Lacy, C.H., Frueh, M.L., and Turner, A.E.} {1986} {Astron.\ J.} {91} {383}

{Prugniel, Ph., and Soubiran, C.} {2001} {Astron.\ Astrophys.} {369} {1048}

{Sandage, A.} {1969} {Astrophys.\ J.} {158} {1115}

{Sandage, A., and Eggen, O.J.} {1959} {MNRAS} {119} {278}

{Spite, M., Francois, P., Nissen, P.E., and Spite, F.} {1996} {Astron.\ Astrophys.} 
{307} {172}

{Taylor, B.J.} {1980} {Astron.\ J.} {85} {242}

{van Leeuwen, F.} {2007} {Astron.\ Astrophys.} {474} {653}

}

\end{document}